# Cation modified $A_2$(Ba, Sr and Ca)$ZnWO_6$ cubic double perovskites: A theoretical study

Rajneesh Chaurasiya[1], Sushil Auluck[2], Ambesh Dixit[1,#]
[1]Department of Physics & Center for Solar Energy, Indian Institute of Technology Jodhpur, 342037, India
[2]National Physical Laboratory, Delhi, India
[#]ambesh@iitj.ac.in

**Abstract:**

The cubic double perovskites $A_2ZnWO_6$ (A = Ba, Sr and Ca) are studied to understand the effect of A cation site, using the density functional theory (DFT) based full potential augmented plane wave method (FP-LAPW) with GGA and mBJ exchange correlation potentials. The structural robustness and stability are investigated using the bond lengths and the total energy. The band structure and density of states suggest that all these cubic double perovskites are indirect wide band gap semiconductors. The band gap varies from 3.90 eV (2.97 eV) for $Ba_2ZnWO_6$ system to 3.40 eV (2.8 eV) for $Ca_2ZnWO_6$ system using mBJ (GGA) exchange correlation potentials. Our studies suggest that cation site modification has a strong effect on physical and electronic properties, in contrast to the structural robustness. The lattice parameter decreases from 8.19 Angstrom to 7.9 Angstrom from Ba to Ca at alkali cation site and the electronic band gap variation follows the common cation rule. The charge densities show enhanced localization of charges near the zinc and oxygen sites with increasing alkali cation atomic radii. In addition, we discuss the impact of A site cation modification on the dielectric and optical properties for $A_2ZnWO_6$ double perovskites.

## Introduction

The normal perovskite $ABO_3$ structures are investigated intensively to understand the numerous physical properties such as intrinsic and relaxor ferroelectrics, piezo/pyro-electrics, and dielectrics. The functionality of these systems can further be engineered by manipulating the cations sites either by different elemental dopants or doubling the lattice sizes with different cations. In conjunction with these ternary perovskites, quaternary double perovskites are gaining attention because of the possibility of manipulating their physical properties either by interplaying cations at different atomic sites or by doping the suitable elements at different cation sites. These double perovskite oxide systems are not only of technological importance but also provide an avenue to understand the physics of the onset of the functional properties. The double perovskites are expressed as $A_2BB'O_6$, which are derived by doubling the normal perovskite $ABO_3$ with different B cations, where A site is occupied by a rare-earth or alkaline-earth metal, B and B' sites are occupied alternately by different cations (transition/non-transition metals) according to their charge and ionic size. Here, the nearest neighbouring interaction is in the form of B–O–B' between different B cation sites in contrast to B-O-B in the normal $ABO_3$ perovskite. These provide a possibility of tailoring the physical properties of the parent compound by suitable substitution or doping at any of the cationic sites [1]. The transition metal based double perovskites have shown promise for multifunctional properties including the simultaneous ordering of magnetism, ferroelectricity, magnetodielectric coupling and electric field controlled magnetic sensors [2]. These interesting properties in double perovskite compounds are observed when one or both of the B site atoms is a transition metal ion, leading to the complex magnetic interactions among different magnetic ion sites. This is attributed to the unpaired d orbital electrons, which are robustly associated with each other through the direct coupling or oxygen 2p electrons mediated coupling through the superexchange interactions [3–11]. The possibility to tune the electronic properties of double perovskite materials poses a materials' design challenge for applications in superconductivity, thermoelectrics, magnetodielectrics and even in solar photovoltaics. The ability to harvest sunlight in the visible spectrum is primarily limited to these double perovskites because of their large optical band gaps. However, there are efforts to minimize the band gap of double perovskite materials by interplaying at A and B cation sites using both experiment and computational approach [12,13]. Berger et al. [13] reported several double perovskites as the probable new low band gap systems, and may be the potential absorber materials to design the all oxide based solar photovoltaic systems, also known as a the next generation in photovoltiacs [14].

The double perovskite oxide systems are getting wider attention due to their potential applications for microwave dielectric resonators, dielectric mediated microwave absorption, filters in mobile phones, capacitors and oscillators etc [15 - 17]. The cubic perovskite systems can be transformed into distorted structures by varying temperature, pressure, chemical composition and electric field [18]. There are also recent reports on relaxor ferroelectrics on cubic perovskite systems and other applications [19 - 30]. Khalyavin et al. [31]showed that $Ba_2MgWO_6$ and $Ba_2ZnWO_6$ ceramic materials exhibit relatively better dielectric properties in the microwave region as compared to other similar double perovskite systems. The rare earth element modified double perovskite systems are investigated for oxide based luminescent materials and have shown promise towards efficient optical materials [32]. The large band gap and relatively large dielectric constant of these materials make them suitable for radiation protection and dosimetery applications [33, 34]. The double perovskite systems mentioned in the present discussion have a cubic structure with Fm-3m space group. Thus, only limited studies on electronic, optical and dielectric properties are available, especially for cubic $A_2ZnWO_6$ (A = Ba, Sr and Ca) double perovskite systems in spite of the huge technological potentials. Recent work by Li et al. [37] showed cation vacancy driven magnetism in non-magnetic double perovskite systems. These systems provide non-zero magnetic moment because of polarized cation vacancy spins, leading to the robust ferromagnetic system, in contrast to the usual oxygen vacancy mediated magnetism in oxides [38]. Considering the same, the present work aims the detailed computational studies on $A_2ZnWO_6$ (A = Ba, Sr and Ca) cubic double perovskite systems to understand the impact of cation modification at A site, especially on structural, electronic, and optical properties at T = 0 K and P = 0 Pa i.e. in ground state. Further, the considered double perovskites are interesting due to their simple cubic crystallographic structure. It would be interesting to understand the crystallographic robustness of such simple cubic systems against A site cation occupancy. We expect that any small modification on atomic sites should lead to the distorted cubic structures, which may give rise to interesting functional properties such as ferroelectricity and piezoelectricity. Hence we thought it useful to investigate the cubic $A_2ZnWO_6$ double perovskite system for A = Ca, Sr and Ba cations. The ab-initio density functional theory (DFT) calculations are carried out and structural and electronic properties such as electronic band structure and density of states are studied intensively. Further, their optical properties have been investigated and discussed including the refractive index, extinction coefficient, reflectivity, absorption spectrum, optical conductivity and electron energy loss spectrum in conjunction with their electronic properties. To draw from our conclusions we found that $A_2ZnWO_6$ systems are quite robust against different A site cations considered in the present study, even with such complex cation/anion combinations. Thus, it becomes important to understand the physical properties of such pristine cubic double perovskite systems.

## Computational details

The first principle calculations are carried out for $A_2ZnWO_6$ (A= Ba, Sr and Ca) double perovskite materials using the full potential linear augmented plane wave (FP-LAPW) method, as implemented in Wein2k [39]. The GGA with Perdew Burke Ernzerhoff type parameterization has been utilized for the exchange-correlation energies within the DFT framework [40]. In addition, modified Becke-Johnson (mBJ) [41] exchange potential based calculations are also performed to understand the electronic dielectric and optical properties of these double perovskite compounds. The basis sets $6s^2$ of Ba, $5s^2$ of Sr, $4s^2$ of Ca, $3d^{10}$, $4s^2$ of Zn, $5d^4$, $6s^2$ of W and $2s^2$, $2p^4$ of O are used for the calculations. The cubic unit cell has been divided into muffin tin radius ($R_{mt}$) around ions and interstitial regions (IR). The values of muffin tin radii for Ba, Sr, Ca, Zn, W, and O are selected appropriately to avoid any atomic sphere overlapping. The plane wave cut off parameters $R_{mt}*K_{max} = 7$ and $G_{max} =12$ are used for optimization and analyzing the various properties of $A_2ZnWO_6$ double perovskite materials. The maximum value of $l$ ($l_{max}$) is considered 12 and cut-off energy has been set at -6 Ry, defining the separation between the core and valence states. In these calculations, relaxation of atomic position has been carried out by minimizing the atomic force 1 mRy/au for each atom. A large plane wave cut-off of 150 Ry is used throughout the calculation and initially 350 k-points are considered in the irreducible Brillouin zone for total energy convergence and the structural properties of these double perovskites. However, 1000 k-points are used to compute the electronic and optical properties for these double perovskites. The self-consistent calculations are carried out for the convergence of the total energy with ~ 0.0001 Ry tolerance.

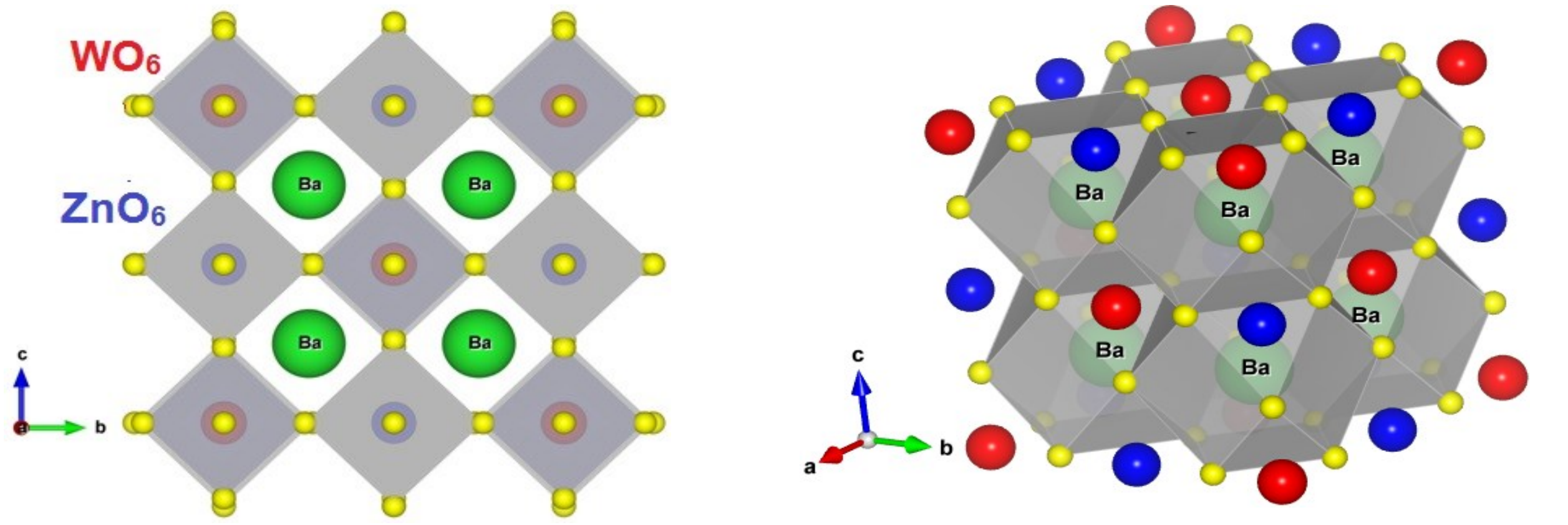

**Figure 1: Crystal structure of $Ba_2ZnWO_6$ cubic double perovskite system, with respective $WO_6$ and $ZnO_6$ octahedron, b-c planer view (left panel) and 3D polyhedral view (right panel)**

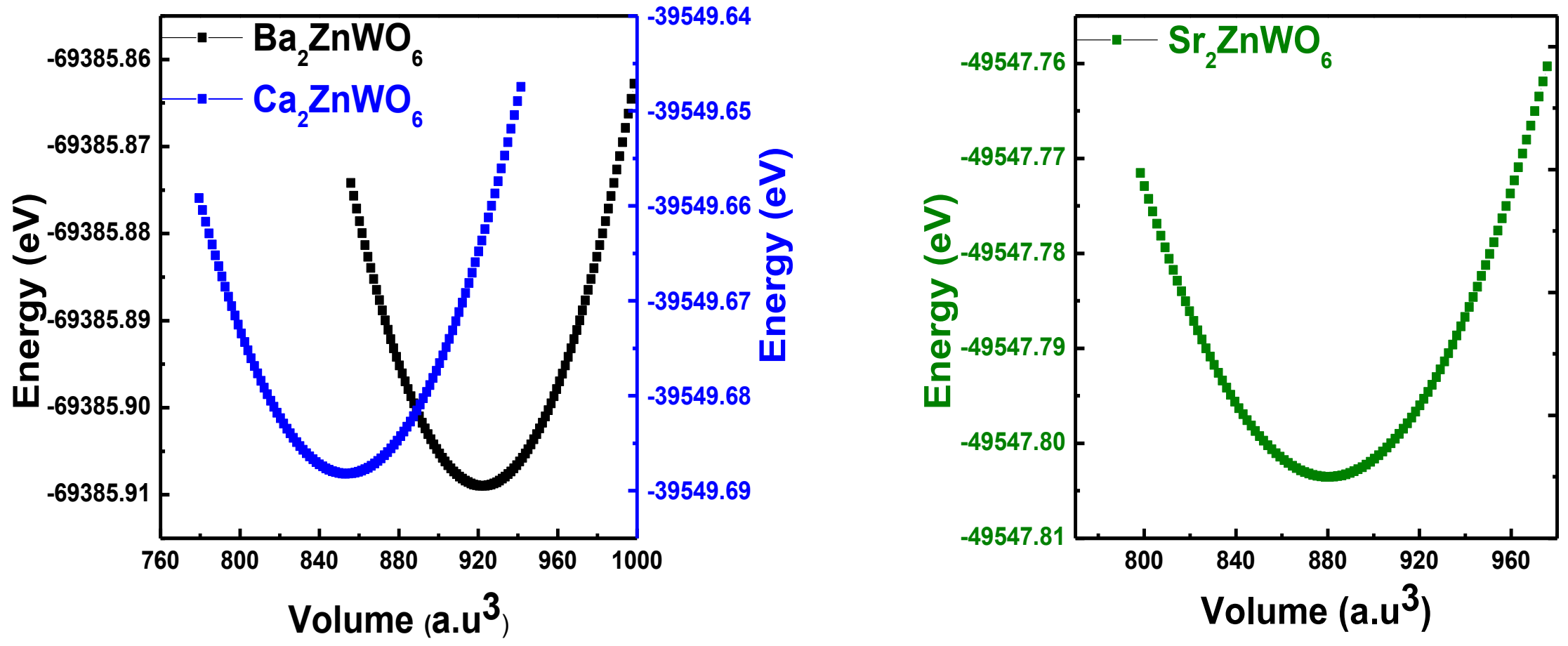

**Figure 2: Total energy as a function of cubic cell volume for $A_2ZnWO_6$ cubic double perovskite systems; left panel for A = Ba and Ca and right panel for A=Sr .**

## Results and Discussion

### Structural properties:

The unit cell of the cubic symmetry (space group Fm-3m) of the double perovskites material is shown in Fig. 1. The Wyckoff positions of the double perovskite are A(0.25, 0.25, 0.25), Zn(0.5,0.5,0.5), W(0,0,0) and O (x,0,0). The position of oxygen atom (x, 0, 0) vaires according to the ionic radii of A site cation, and the optimized values of x are 0.237, 0.241 and 0.245 for Ba, Sr, and Ca cations in $A_2ZnWO_6$ double perovskites respectively. The representative structure of optimized cubic double perovskite ($Ba_2ZnWO_6$) is shown in Fig.1, where Zn and W cations are octahedrally coordinated with the six oxygen atoms, A site cations are coordinated with the twelve oxygen atoms, which form a polyhedral network and each Zn cation has six W cations in the nearest neighbour positions and vice versa. The bond lengths of $WO_6$ and $ZnO_6$ decreases with the size of A site cation and are summarized in Fig.4. The W-O bond length is always larger than that of Zn-O except for $Ba_2ZnWO_6$ where Zn-O bond length is larger than W-O bond length. The lattice parameters are optimized by varying the volume of the cubic cell and the observed results are simulated using the Murnaghan's equation of state [42], Eq.1. The internal parameter x has been optimized such that the forces are 1mRy/au.

$$E(V) = E_0 + \frac{B_0 V}{B_0'}\left[\frac{\left(\frac{V_0}{V}\right)^{B_0'}}{B_0'-1} + 1\right] - \frac{B_0 V}{B_0'-1} \quad \text{---------------------------} \quad (1)$$

where $E_0$ is the minimum energy corresponding to the equilibrium volume $V_0$ at temperature T=0 K, $B_0$and $B_0'$ are the bulk modulus and first order derivative of the bulk modulus with respect to pressure, respectively at the equilibrium volume. The total energy versus volume curves are shown in Fig. 2 for all the considered double perovskites considered in this work. The fitted Murnaghan's equation has been used to estimate the bulk modulus values for these double perovskites. The optimized lattice constants are 8.19 Å, 8.02 Å and 7.90 Å for $Ba_2ZnWO_6$, $Sr_2ZWO_6$, and $Ca_2ZnWO_6$ double perovskites, respectively, using GGA exchange correlation potential, and are listed in Table1. The variation of these constants against different cations is plotted in Fig.3. The calculated $Ba_2ZnWO_6$ and $Sr_2ZnWO_6$ lattice parameters are in agreement with the available experimental data listed in Table 1. The lattice parameter decreases with decreasing ionic radii from Ba to Ca for the electropositive A site cation. The ionic radius of A-site cation is greater than that of B-site cation, and has a strong dependence on lattice parameter. The other physical parameters, such as equilibrium volume, bulk modulus and band gap, are summarized in Table 1 for these double perovskites, showing the best match with the available experimental and computational results.

**Table 1: Lattice parameters, equilibrium volume, bulk modulus and band gap of $A_2ZnWO_6$ double perovskite systems using GGA and mBJ exchange potentials are compared with the available literature reports**

| AZWO | Lattice constant ( Å) | Volume (a.u.$^3$) | Bulk modulus (GPa) | Band gap Eg(eV) | | | Reference |
|---|---|---|---|---|---|---|---|
| | | | | GGA | mBJ | Experimental | |

| | Present | Theoretical Reported | Experimental Reported | Present | Reported | Present | Reported | Present | Reported | Present | Reported | Reported | |
|---|---|---|---|---|---|---|---|---|---|---|---|---|---|
| $Ba_2ZnWO_6$ | 8.19 | 8.11[a], 8.09[d] | 8.14[c], 8.12[f] | 929 | | 167.95 | | 2.97 (L-X) | 2.99[a,d] | 3.90 | 3.92[d] | 3.50[b] | [a]Ref. [10] [b]Ref. [12] |
| $Sr_2ZnWO_6$ | 8.02 | | 7.96[e] | 872 | | 158.89 | | 2.90 (L-X) | | 3.80 | | | [c]Ref. [22] [d]Ref. [33] |
| $Ca_2ZnWO_6$ | 7.90 | | | 832 | | 221.41 | | 2.80 (L-X) | | 3.40 | | | [e]Ref. [34] [f]Ref. [42] |

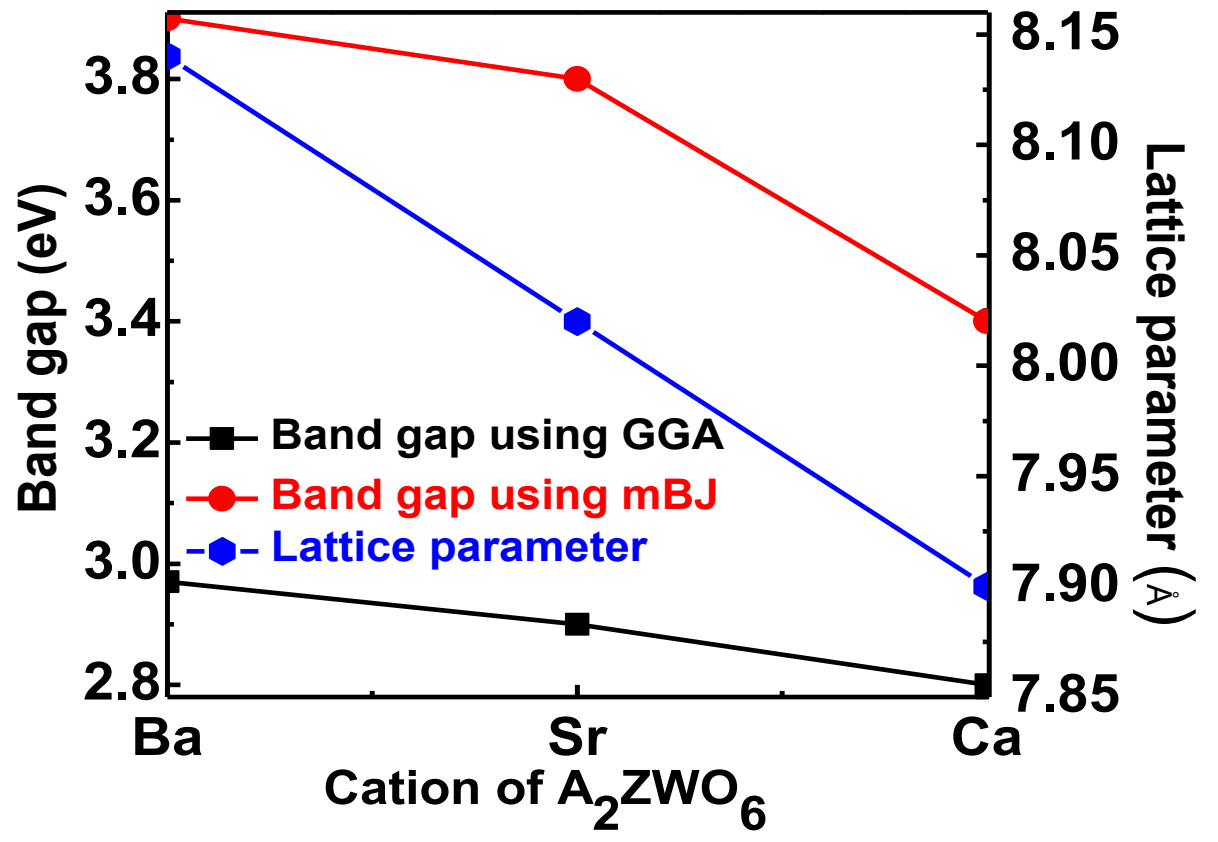

Figure 3: Effect of cation on the lattice parameter and band gap for $A_2ZnWO_6$ double perovskite systems

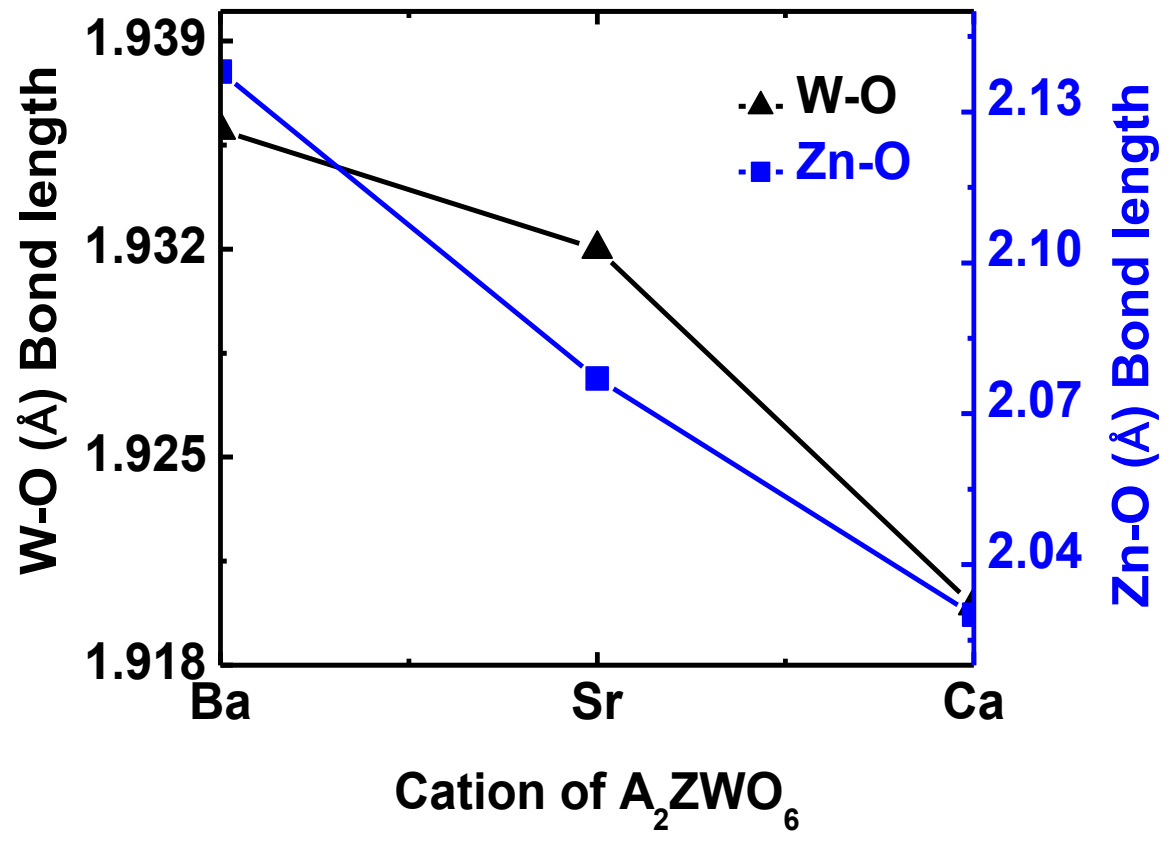

Figure 4: Effect of cation on bond length of $ZnO_6$ and $WO_6$ octahedral symmetry in $A_2ZnWO_6$ double perovskite systems

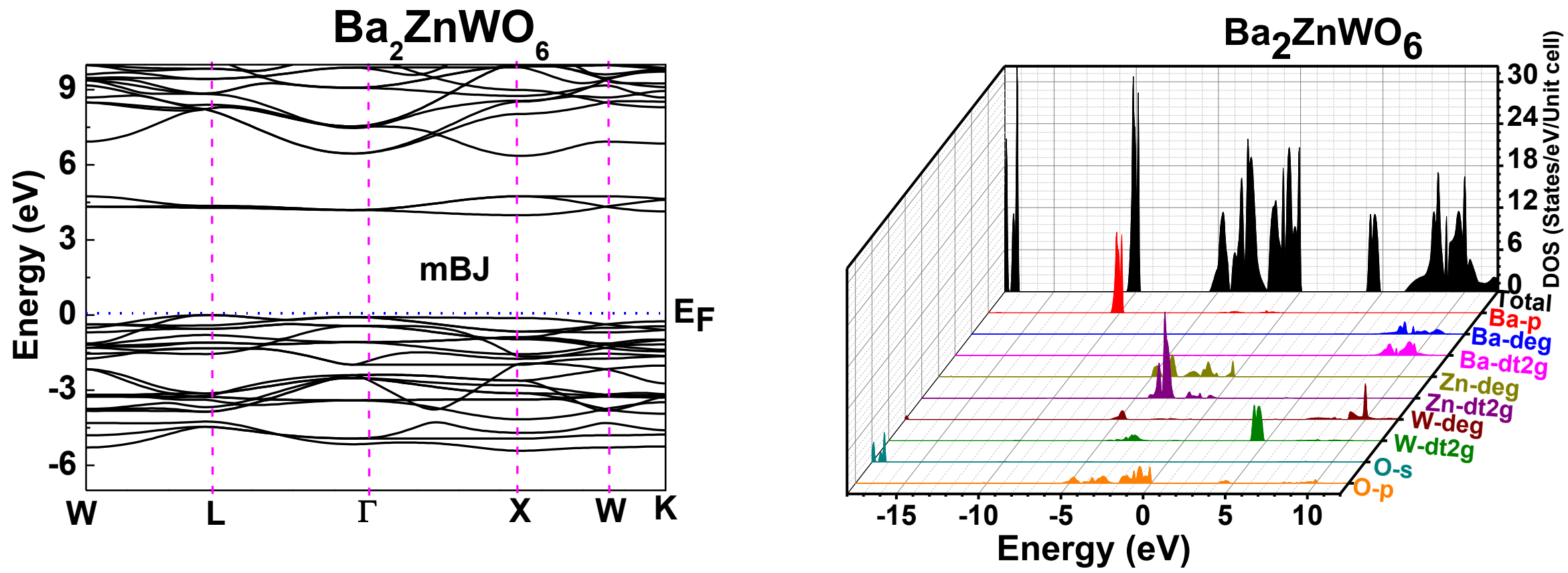

Figure 5: Electronic band structure (left panel) and total and projected density of states (right

**panel) of $Ba_2ZnWO_6$ double perovskite system**

## Electronic properties

The electronic properties of the double perovskites are calculated for the optimized structure using both GGA and mBJ exchange correlation potentials. The mBJ calculations are relatively better for energy band gaps estimation and results are in close agreement with the experimental observations, we present mBJ results only and mentioned GGA results, wherever necessary. The electronic band structures are computed along W-L-Γ-X-W-K symmetry points of the reciprocal lattice. The Fermi energy is set at 0 eV and the computed $Ba_2ZnWO_6$ electronic band structure is shown in Fig. 5 (left panel) for mBJ exchange correlation potential. The valence band maxima (VBM) is located at L symmetry point and the conduction band minima (CBM) is located at X symmetry point, suggesting an indirect band gap for $Ba_2ZnWO_6$ with ~ 2.97eV (Γ-X) band gap using GGA-PBE exchange correlation potential. This is in agreement with the work of Sahnoun et al. [10], but less than the experimentally observed band gap by Enget al. [12]and Day et al. [43]. This is attributed to the underestimation of energy gaps using GGA-PBE exchange correlation potentials. In contrast mBJ based calculations are consistent with an indirect band gap ~ 3.90 eV for $Ba_2ZnWO_6$ system in the present study, slightly. This value is slightly higher than 3.5 eV, the reported experimental value [23] but closer to that of the calculated by Dong et al.~ 3.92 eV [34] using mBJ. The bottom of the conduction band is nearly flat along W-L-Γ. This flatness suggests a large carrier effective mass, which is not changing significantly along W-L-Γ. However, there is large variation along high Γ-X and X-W-K suggesting that the effective mass may vary along these lines.

The electronic band structure of $Sr_2ZnWO_6$ is plotted in Fig.6 (left panel) for mBJ exchange correlation potential. The valance band minimum (VBM) and conduction band minimum (CBM) points are located at L and X symmetry points, confirming an indirect band gap of ~ 2.90eV using GGA and ~ 3.8eV using of mBJ exchange correlations. A similar band dispersion has been seen for $Ca_2ZnWO_6$ (Fig. 7 left panel), with a band gap ~ 3.4 eV using mBJ exchange correlation potential. The minimum of the conduction band is nearly flat along L-Γ-X but the small dispersions are noticed along W-L and X-W-K for both $Sr_2ZnWO_6$and $Ca_2ZnWO_6$. Similarly, the top of the valance band is exhibiting nearly linear band dispersion along L- Γ symmetry points in conjunction with a small dispersion along other Brillion zone points.

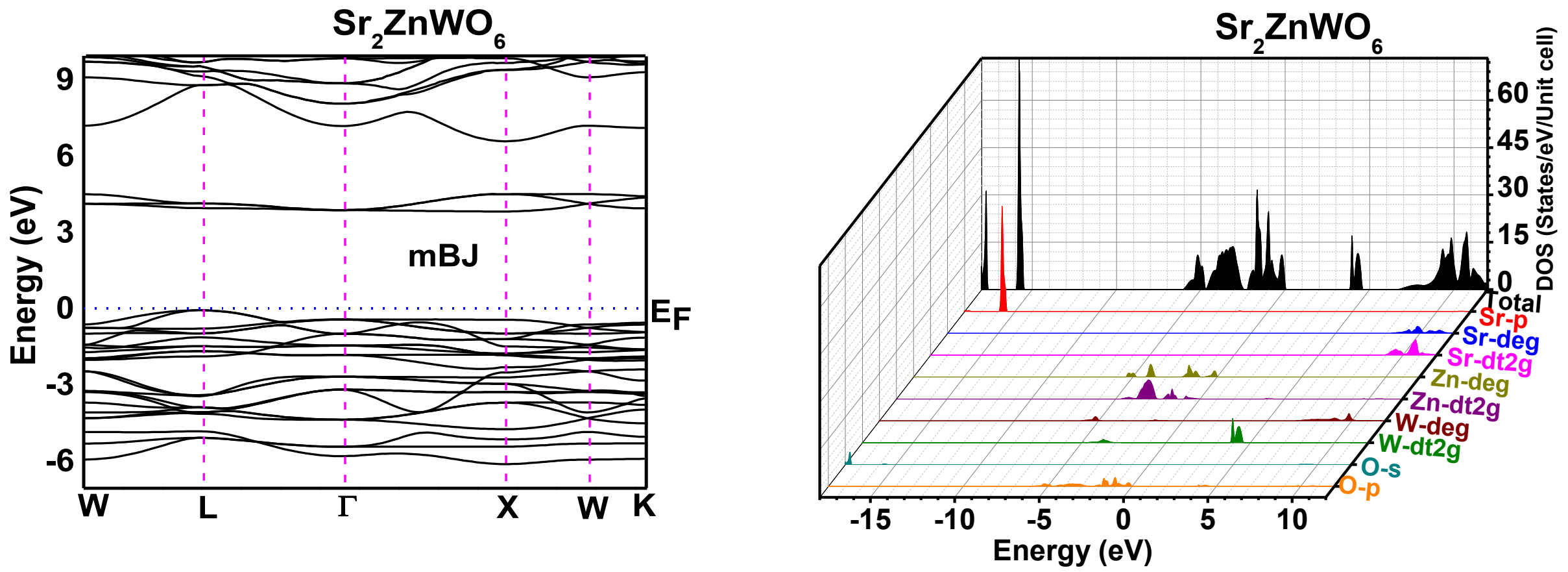

**Figure 6: Electronic band structure (left panel) and total with projected density of states (right panel) of $Sr_2ZnWO_6$ double perovskite system**

The total density of states (DOS) and projected density of states (PDOS) are computed using the GGA and mBJ exchange correlation potentials. The B cation sites form an octahedral with the oxygen atoms in $A_2B_1B_2O_6$ double perovskites. The electronic states of B cation loses their degeneracy in such octahedral configuration and splits into a doubly degenerate $e_g$ higher energy and triply degenerate $t_{2g}$ lower energy states, respectively. The observed splitting is attributed to crystal field associated with $WO_6$ and $ZnO_6$ octahedrons. The DOS and PDOS spectra of $Ba_2ZnWO_6$ are shown in Fig. 5(right panel), indicating that the occupied p states of oxygen atoms contribute to the valence band and conduction band, consisting of the triply degenerate ($t_{2g}$) states of tungsten effectively. A similar DOS result for $Ba_2ZnWO_6$has been reported by the Sahnoun et al. [10] and Dong et al. [34]. The PDOS plots show the contribution of oxygen and tungsten atoms s states in the deep valance band, attributed to their occupied nature. VBM near the Fermi energy (0 to -5eV) are mostly dominated by the oxygen p states and around -3.0 eV a strong hybridization can be seen between the oxygen p and zinc d-$t_{2g}$ orbitals. Moreover, d-$t_{2g}$ orbital of zinc and tungsten show localization at -3.0 eV and 4.5eV respectively. The completely filled zinc d states contribute to the formation of the valance band in the 0 to -5.0 eV energy range. However, the CBM consists of tungsten d-$t_{2g}$ states, which are hybridized with oxygen p states near 4.5eV. The density of states in 6.5-11eV are mostly dominated by tungsten d-$e_g$ states and a little contribution from barium d and oxygen p states.

The total and partial densities of states are shown in Fig. 6 (right panel) for $Sr_2ZnWO_6$ system. The VBM consists of the doubly degenerated d-$e_g$ zinc states and oxygen p states, showing strong hybridization near the Fermi energy. The valance band of $Sr_2ZnWO_6$ has contributions mainly from the triply degenerate zinc d-$t_{2g}$ atomic states, which are more localized as compare to the d-$e_g$ states. In addition, zinc d-$e_g$ states are delocalized from -5.0 eV to the Fermi energy in the valance band and delocalization is maximum near the -5.0 eV. The oxygen p orbitals also show the delocalization in the range of -6.0 eV to Fermi energy. Sr, Zn, W cations and O anion s orbital states lie in the semi-core region and hence, do not play any significant role near the Fermi energy. The conduction band minima (CBM) of $Sr_2ZnWO_6$ **(**at ~ 3.8eV using mBJ calculation) consists of tungsten d-$t_{2g}$ states and its contribution is attributed to its' partially filled d states. Similarly the DOSs and PDOSs spectra of $Ca_2ZnWO_6$are plotted in Fig.7(right panel) using mBJ calculation, substantiating the observed band gap value ~ 3.4eV, as shown in the respective electronic band dispersion. In this case, the formation of valance band is due to the zinc d$e_g$ states and oxygen p states. Here also, a strong hybridization has been observed between these states near the Fermi energy. The atomic configurations of Zn and W atoms are [Ar] 3d10 4s2 and [Xe] 4f14 5d4 6s2, showing that the s states of zinc and tungsten lie in the semi core region, that's why the contribution of respective s states is not significant near the Fermi energy. The tungsten d$t_{2g}$ states contribute mainly in the formation of conduction band together with a minor contribution from oxygen p states. The states near -3.0 to -5.0 eV in DOSs are due to the oxygen p, zinc d$e_g$ and d$t_{2g}$, and tungsten d$t_{2g}$ states.

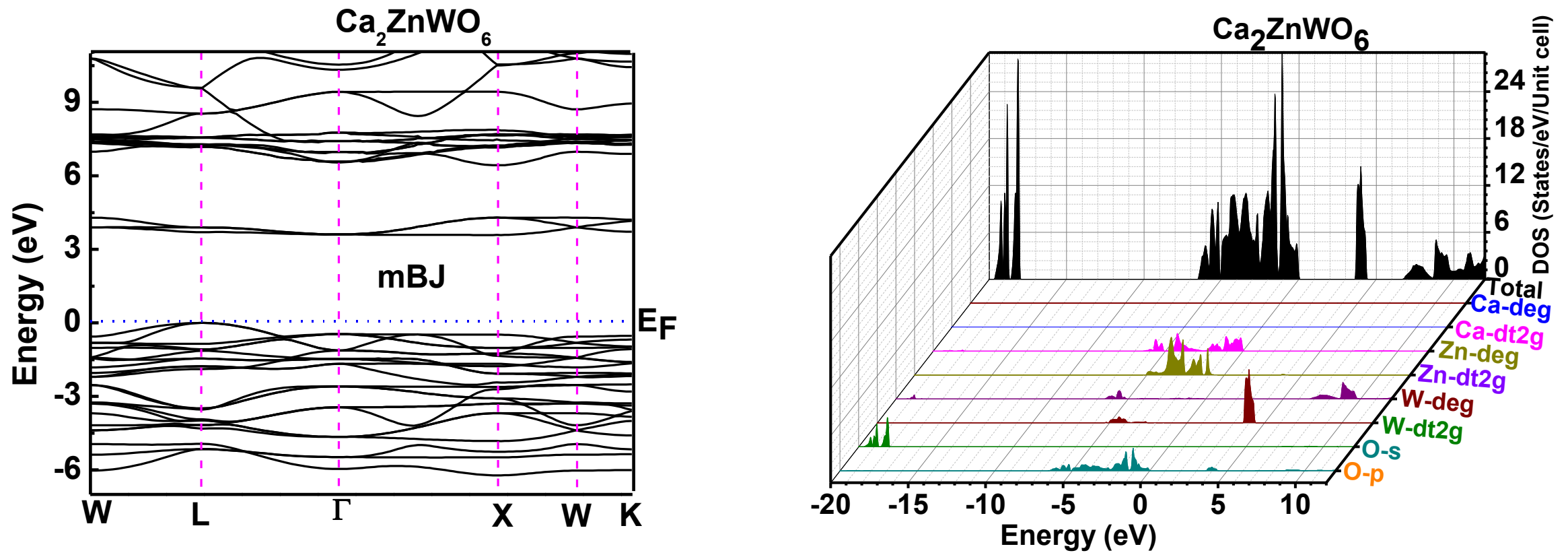

**Figure 7: Electronic band structure (left panel) and total and projecteddensity of states (right panel) of $Ca_2ZnWO_6$ double perovskite system**

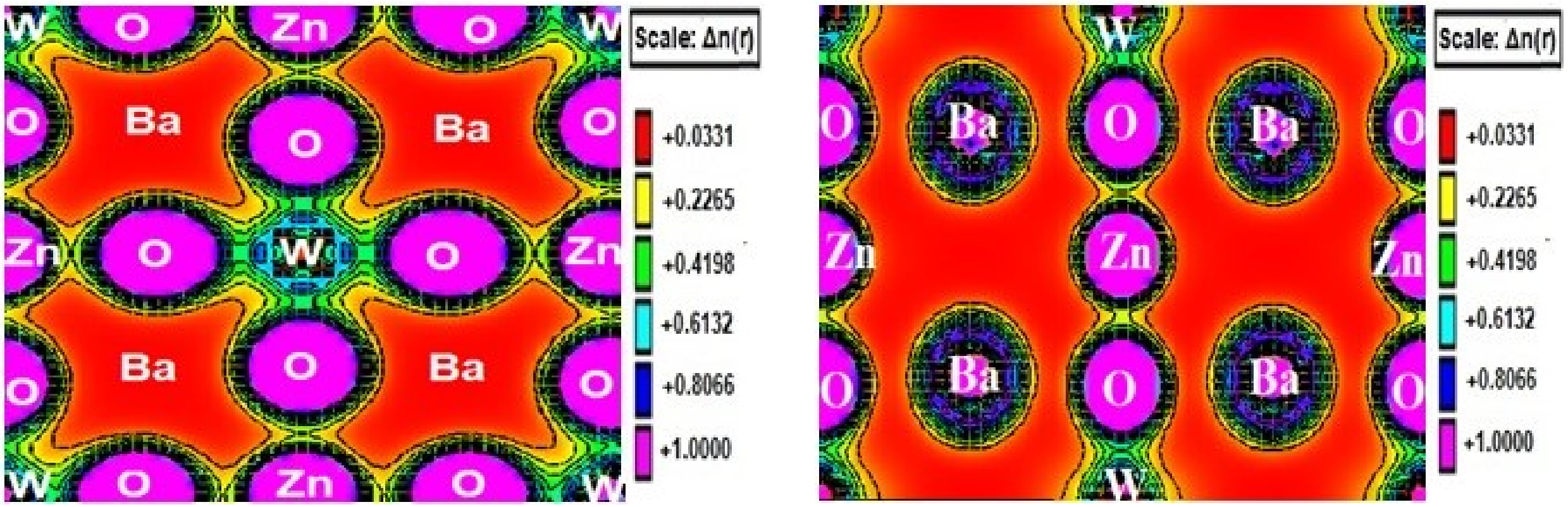

**Figure 8: Valance electron charge density contour plot of $Ba_2ZnWO_6$ using GGA-PBE for (100) plane (left panel)and (110) plane(right panel)**

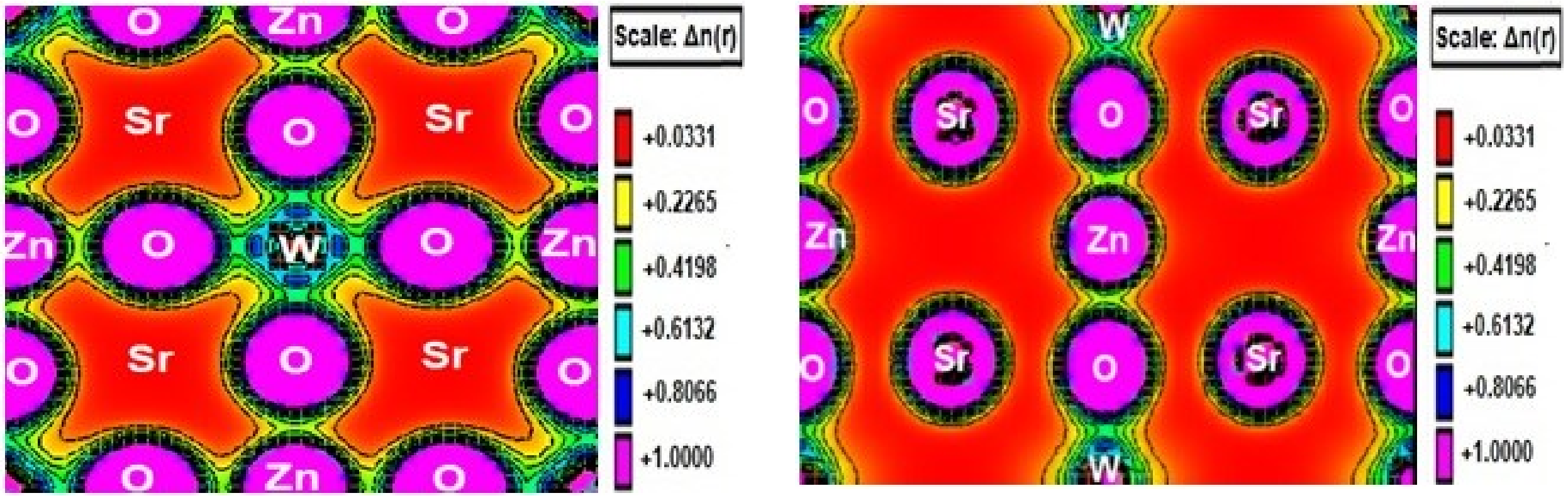

**Figure 9: Valance electron charge density contour of $Sr_2ZnWO_6$ using GGA-PBE for (100) plane (left panel) and (110) plane (right panel)**

**Charge density analysis:**

The behaviour of bonding and charge sharing between the atoms can be understood using the valance electron charge density derived from the converged wave functions. The valence electron charge densities are computed for (100) and (110) crystallographic planes and results are summarized in Fig. 8–10. The charge density plots show the maximum density near the oxygen atom because of its large electronegative nature for all these double perovskite compounds, which can be easily seen in the respective color scales, where pink color (+1.0) corresponds to the maximum charge accumulating site. The change density plot for $Ba_2ZnWO_6$ along (100) plane is shown in Fig.9(left panel), showing the sharing of charge between the W-O and Zn-O atoms because of $WO_6$ and $ZnO_6$ octahedral symmetry. This charge sharing among these atoms confirms the formation of covalent bonds. This charge sharing between W-O and Zn-O also has also been verified in the PDOS spectra where strong hybridization has been seen between the atomic states of these atoms, as explained earlier. A similar nature has also been observed for $Sr_2ZnWO_6$ and $Ca_2ZnWO_6$ as can be seen from the planar charge densities shown in Fig. 9 and 10 respectively.

We have plotted the charge densities along (110) plane to understand the contribution of A site cation with other atoms in these double perovskite systems. The A site cations are isolated from the rest of atoms in (110) plane, substantiating the ionic bonding nature with the oxygen. The localization of charge due to A site cation decreases in the order from Ba – Sr – Ca, as shown in the respective charge densities of $Ba_2ZnWO_6$ (Fig. 8), $Sr_2ZnWO_6$ (Fig. 9) and $Ca_2ZnWO_6$ (Fig. 10). This is attributed to the enhancement in the electro-negativity and simultaneously decrease in the ionic radii of A site cations from Ba – Sr – Ca. According to Pauling electronegativity difference scale, Ba (0.89), Sr (0.95), Ca (1.0), W (2.36), Zn (1.65) and O (3.44) differ in the charge density plot substantiating the polar covalent bonding nature between the Zn-O and W-O, and the ionic bonding nature between A site cations and oxygen atom. The relatively small electronegativity difference among Zn and tungsten with oxygen is attributed to the observed charge sharing and covalent bonding for Zn-O and W-O atoms in all these compounds. The large difference in electronegativity among A site cations and oxygen is responsible for the respective charge transfer among different atoms, resulting into the ionic bonding between A site cations and oxygen.

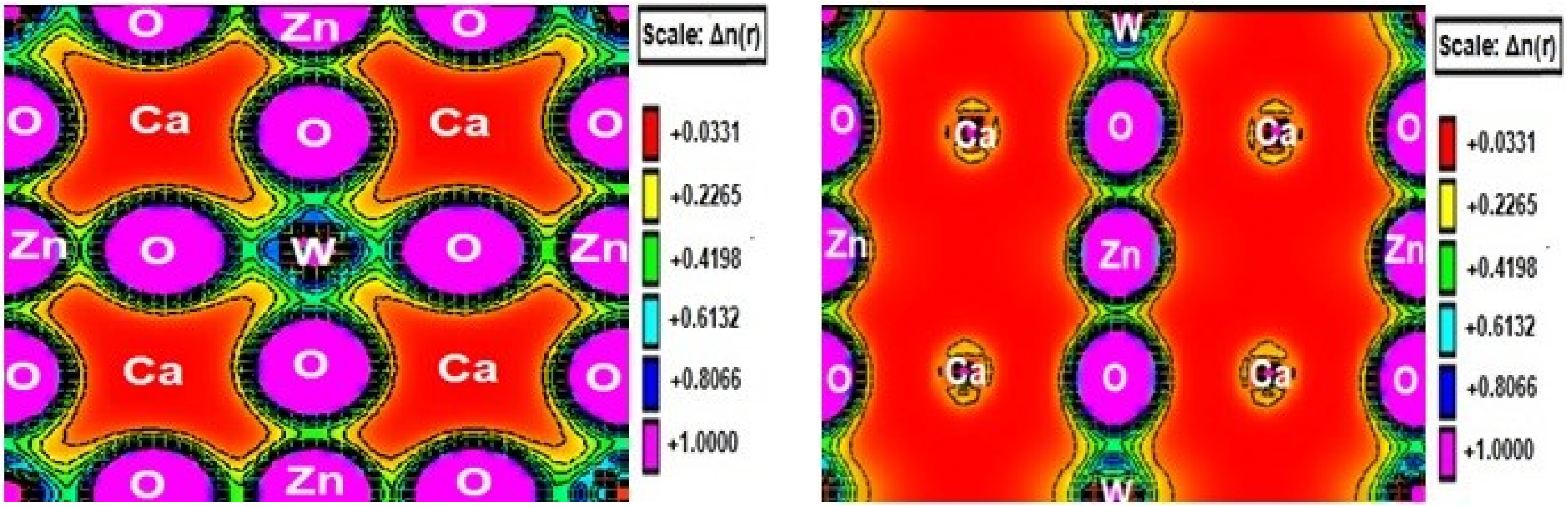

**Figure 10: Valance electron charge density contour of $Ca_2ZnWO_6$ using GGA-PBE for (100) plane (left panel) and (110) plane (right panel)**

## Optical properties

The optical properties can be calculated from the band structure properties.$A_2ZnWO_6$ double perovskite systems are cubic in nature and hence should exhibit isotropic optical properties. The calculated imaginary and real dielectric constants as a function of energy using mBJ potentials are plotted in Fig. 11(left panel) and 11(right panel) respectively for different double perovskite compounds. The imaginary dielectric constant is vanishingly small (nearly zero) upto certain energy, Fig. 11(left panel), and the onset of non-vanishing imaginary dielectric constant at a finite energy is consistent with the calculated band structures, as discussed earlier.

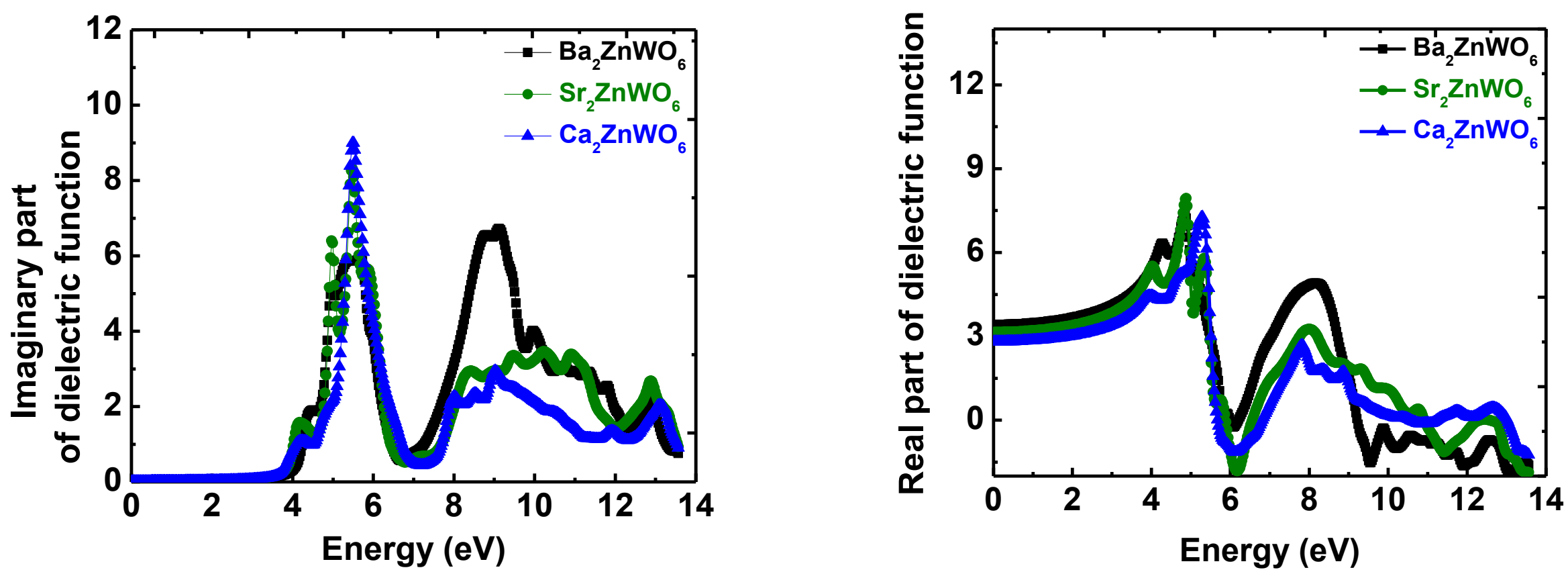

**Figure 11: Dielectric function of $A_2ZnWO_6$: imaginary dielectric function (left panel) and real dielectric function (right panel), calculated using mBJ exchange potentials**

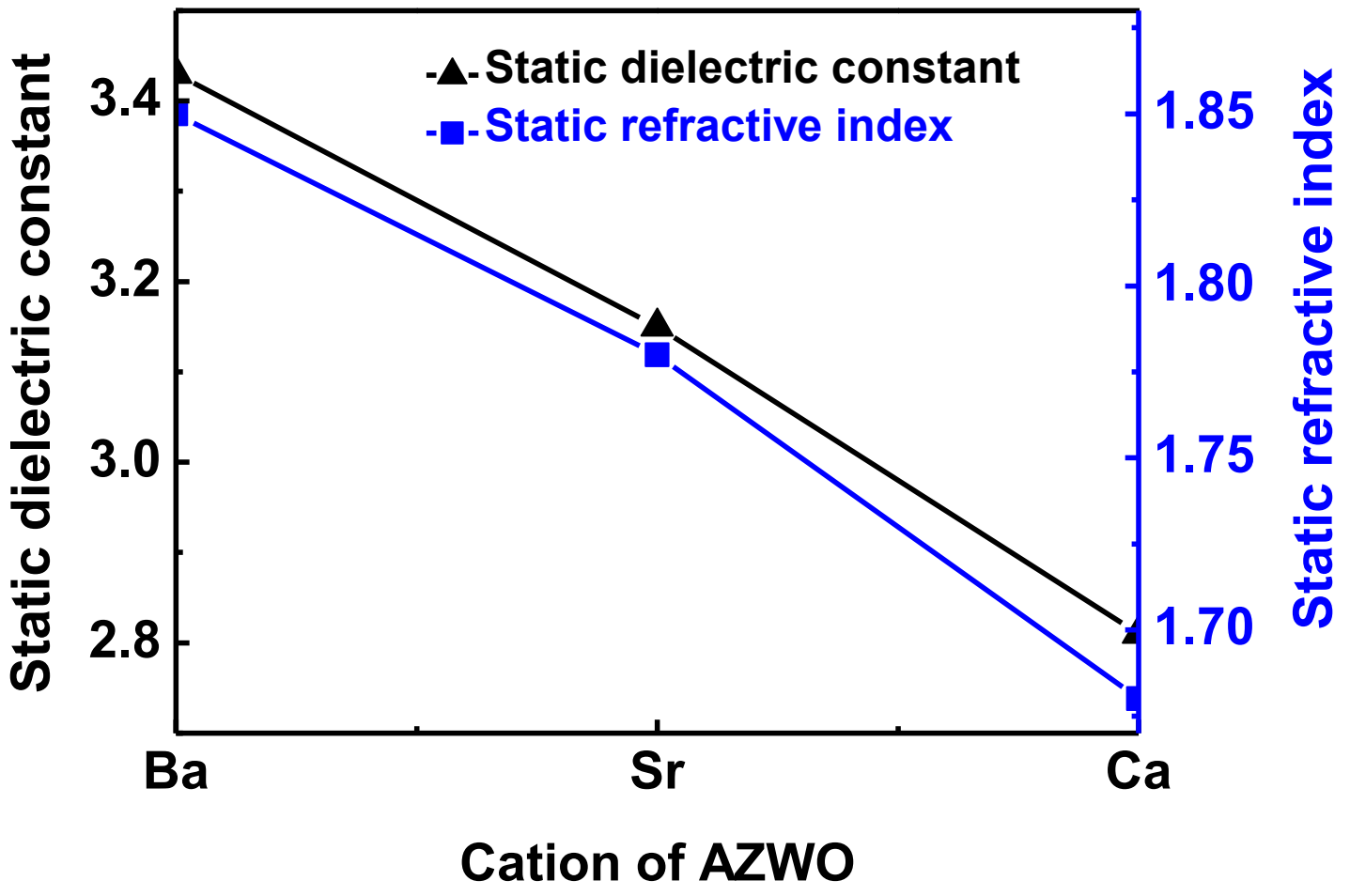

**Figure 12: Effect of cation at A site in $A_2ZnWO_6$double perovskite on the static dielectric constant and static refractive index using mBJ exchange correlation potentials**

The peaks in the imaginary part of dielectric function reveal the interband transition from the valance bands to the respective conduction bands. The respective peaks in the imaginary dielectric constant for $Ba_2ZnWO_6$, $Sr_2ZnWO_6$ and $Ca_2ZnWO_6$ can clearly be identified in Fig. 11(left panel).The electronic band structure and DOS spectra of $A_2ZnWO_6$ systems suggest that multiple transitions can be realized depending on the incident photon energy. The spectral variation of $Ba_2ZnWO_6$ is consistent with Dong et al. [34] computational work. The main peaks in the imaginary part of dielectric function are found at 4.5eV, 4.17eV and 4.1eV for $Ba_2ZnWO_6$, $Sr_2ZnWO_6$ and $Ca_2ZnWO_6$ respectively, Fig.11 (left panel). The other relevant peaks are attributed to the high energy interband transition, as shown in Fig. 11(left panel). The spectral dependence of dielectric function is similar for $Ba_2ZnWO_6$, $Sr_2ZnWO_6$ and $Ca_2ZnWO_6$ because of their similar electronic configuration e.g. Ba ([Xe]$6s^2$), Sr ([Kr]$5s^2$) and Ca ([Ar]$4s^2$). The PDOS pattern is nearly similar for these compounds; however, there are relatively small differences in energy, which finally lead to small energy shifts in dielectric properties [44]. The PDOS of Zn-d and O-p states are hybridized in the valance band maxima and W-5d states are appearing in the conduction band minima, as evident from Fig. 5, 6 and 7 (right panel). The hybridization of atomic orbital will lead to the more intense peaks [45]in dielectric and optical properties and the observed hybridization has been attributed for the same. The peaks, observed in the imaginary dielectric function in the range of 0-7 eV, arise due totransitions form the valance band O-2p, Zn-3d electronic statesto the conduction band W-5d electronic states. The higher energy peak may correspond to transition from semi-core electron of valance band to the conduction band.

The real part of the dielectric function was computed from imaginary part of dielectric function using Kramers-Kronig relation for these double perovskite materials, and results are plotted in Fig.11 (right panel). The computed static dielectric constant $\varepsilon_1(0)$ using mBJ exchange correlation potential without any lattice vibrations are 3.43, 3.15 and 2.81 and showing the maxima at 4.28eV, 4.03 and 4eV for $Ba_2ZnWO_6$, $Sr_2ZnWO_6$ and $Ca_2ZnWO_6$ respectively. The computed values of static dielectric constants using the mBJ approximation are less than that of the GGA approximation, shown in Fig.12 and can be understood using Penn model [46] that explains the inverse relation between the semiconductor band gap and static dielectric constant as $\varepsilon_s(0) = 1 + \left(\frac{\hbar\omega_p}{E_g}\right)^2$; where $\omega_p$ is the plasmon frequency and $E_g$ is the band gap.

Thus, the real part of dielectric function also substantiates the observed electronic structures of $A_2ZnWO_6$ double perovskite systems. Further, a sharp decrease has been observed with increasing the photon energy, from 5.5 -6.5eV, suggesting that the minimal energy – matter interaction in this particular energy range. Further, a zero cross-over from positive to negative real dielectric constant can be seen, Fig. 11 (right panel) at higher energies. Such cross-over suggests the transition into metallic states at higher energies. This may correspond to the screened plasma frequency and may lead to the enhanced reflectivity in this energy region, as seen in Figure 13(right panel) and also can be inferred from reflectivity plots, Fig. 14 (left panel).

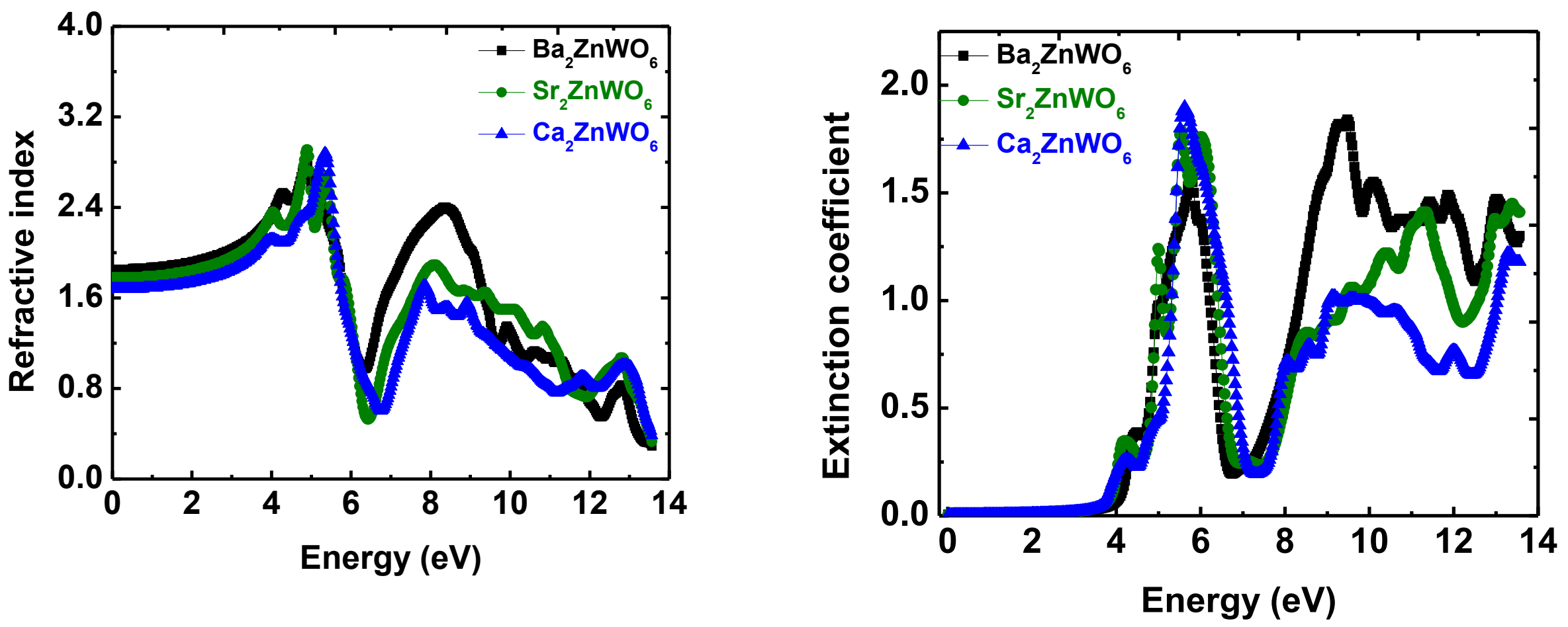

**Figure 13: Refractive index and extinction coefficient of $A_2ZnWO_6$ systems: refractive index (left panel) and extinction coefficient (right panel), calculated using mBJ exchange potentials**

The refractive index and extinction coefficients are computed using both real and imaginary dielectrics in the energy range 0-14eV, and results are displayed in Fig.13 (left panel) and 13(right panel) respectively. The static refractive index can be expressed as $\sqrt{\varepsilon_s(0)}$; and the respective values are 1.85, 1.78 and 1.68 for $Ba_2ZnWO_6$, $Sr_2ZnWO_6$, $Ca_2ZnWO_6$ and $Mg_2ZnWO_6$ respectively. The static dielectric constants, also computed using GGA, not shown here, follow the similar trend as calculated using mBJ approximation, Fig.12. In the lower energy range, refractive index is insensitive to the frequency and exhibits increase with the photon energy but after certain photon energy, it starts decreasing, which may be due to the transitions among different electronic states. The maximum value of refractive index has been observed at same point as the real part of dielectric function. The qualitative natures of refractive indices and extinction coefficients are similar to that of the real and imaginary parts of the dielectric function.

Reflectivity is another important parameter to understand the optical properties of a material, which may directly provide the information about reflection and transmission properties of the materials, useful for numerous applications such as spectrally selective coatings for solar thermal applications, transparent conducting oxides for transparent electronics applications [47]. The reflectivity has been calculated for these double perovskite systems and values are very small, less than 20% under GGA and even lesser ~10% under mBJ for all double perovskite samples, Fig. 14(left panel). The deep valley region (4.5eV-6.5eV) in the reflectivity curve can be understood in terms of the maximum absorption near the electronic band gap of these materials. At higher energies, these materials may experience large electronic transition from valence to conduction band and thus may lead to high reflectivity in such high energy regions, where these materials may experience a transition from the semiconductor to the metallic nature. This is in agreement with the observed dielectric dispersion in 10-14eV energy range.

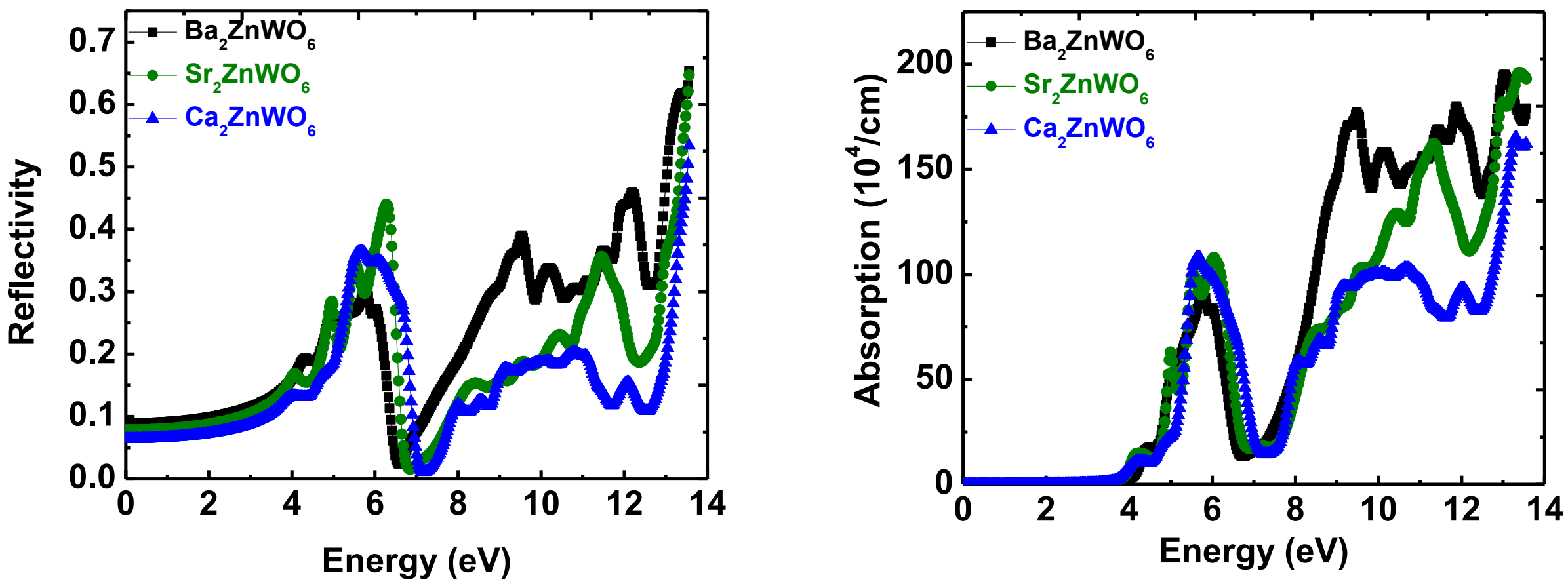

**Figure 14: Reflectivity (left panel) and absorption (right panel) spectra of $A_2ZnWO_6$systems calculated using mBJ exchange potentials**

The optical absorption spectrum of these double perovskite materials as a function of photon energy are shown in Fig.14 (right panel). The corresponding extinction coefficients are plotted in Fig. 13 (right panel) for these systems. The nearly zero absorption and extinction coefficients below band gap energies are accompanied by the interband transition, giving rise to the interband absorption in these double perovskite systems. This implies that above the threshold i.e. band gap energies, electrons can gain sufficient energy to fall into the conduction band. The large band gap absorption in these systems may be useful to develop the highly sensitive ultraviolet sensors and also in thin film geometries, these materials systems can be used for ultraviolet filter applications. In contrast to these relatively large band gap double perovskite, $Ca_2ZnWO_6$ system showed the smaller band gap ~3.4 eV because of transition between oxygen p electrons and tungsten $dt_{2g}$ states electron. The relatively low band gap of $Ca_2ZnWO_6$ system may be useful for several optical and light induced catalytic applications.

## Conclusion

We investigated the impact of A site cation on structural stability, electronic and optical properties of $A_2ZnWO_6$ double perovskite using the density functional theory. The structural stability of $Ba_2ZnWO_6$ is in agreement with the available computational and experimental results. The band gap of $A_2ZnWO_6$ perovskite investigated using the GGA-PBE and mBJ type exchange potentials show strong A site cation dependence with a relatively lower band gap ~3.4 eV for $Ca_2ZnWO_6$ double perovskite among the investigated systems. These systems exhibit the covalent bonding nature among Zn-O and W-O atomic sites with small ionic contributions among A-O atomic sites, as observed from charge density studies. The DOSs and PDOSs explain the contribution of different atomic states towards forming the electronic band structure. The optical properties have been investigated using complex function in terms of refractive index, extinction coefficient, reflectivity absorption, optical conductivity and electron energy loss analysis. The findings may provide the insights for probable applications such as ultraviolet sensors for solar blind regions, photovoltaic electrode materials for dye/quantum dot sensitized solar cells, radiation hard electronics for crystallographically robust $A_2ZnWO_6$ (A = Ba, Sr and Ca) double perovskite materials.

**Acknowledgement:**

Author Ambesh Dixit acknowledges the financial assistance from the Department of Science and Technology (DST), Government of India through the project # DST/INT/ISR/P-12/2014.SA would like to acknowledge the use of the High Performance Computing (HPC) facilities at Physics Department of Indian Institute of Technology in Kanpur (IITK), Intra-University Accelerator Centre (IUAC) in New Delhi, Institute of Mathematical Sciences (IMSC)in Chennai, Council of Scientific and Industrial Research Fourth Paradigm Institute (CSIR-4PI) at Bangaluru and University of Hyderabad in Hyderabad.